# SNe Ia Redshift in a Non-Adiabatic Universe

**Rajendra P. Gupta**[1,2,*]

[1] Macronix Research Corporation, 9 Veery Lane, Ottawa, Canada  K1J 8X4
[2] National Research Council, Ottawa, Canada (retired)
[*] Correspondence: guptarp@macronix.ca

**Absract:** By relaxing the constraint of adiabatic universe used in most cosmological models, we have shown that the new approach provides a better fit to the supernovae Ia redshift data with a single parameter, the Hubble constant $H_0$, than the standard $\Lambda$CDM model with two parameters, $H_0$ and the cosmological constant $\Lambda$ related density $\Omega_\Lambda$.  The new approach is compliant with the cosmological principle.  It yields the $H_0 = 68.28\ (\pm 0.53)$ km s$^{-1}$Mpc$^{-1}$ with an analytical value of the deceleration parameter $q_0 = -0.4$.  The analysis presented is for a matter only, flat universe.  The cosmological constant $\Lambda$ may thus be considered as a manifestation of a non-adiabatic universe that is treated as an adiabatic universe.

____________________________________________________________________________________

1. **Introduction**

   The redshift of the extragalactic objects, such as supernovae Ia (SNe Ia) is arguably the most important of all cosmic observations that are used for modeling the universe.  Two major explanations of the redshift are the tired light effect in the steady state theory and the expansion of the universe [1].  However, since the discovery of the microwave background radiation by Penzias and Wilson in 1964 [2], the acceptable explanation for the redshift by mainstream cosmologist has steadily shifted in favour of the big-bang expansion of the universe, and today alternative approaches for explaining the redshift are not acceptable by most cosmologists.  The situation has been most succinctly expressed by Vishwakarma and Narlikar in a recent paper [3] as follows:  "… a recent trend in the analysis of SNeIa data departs from the standard practice of executing a quantitative assessment of a cosmological theory—the expected primary goal of the observations [4,5]. Instead of using the data to directly test the considered model, the new procedure tacitly assumes that the model gives a good fit to the data, and limits itself to estimating the confidence intervals for the parameters of the model and their internal errors. The important purpose of testing a cosmological theory is thereby vitiated."

   Interestingly, it is the close analysis of the cosmic microwave background that has created tension between the Hubble constant derived from the spectral data and from the microwave background data [6, 7].

   The status of the expanding universe and steady state theories has been recently reviewed by López-Corredoira [8] and Orlov and Raikov [9].  They concluded that based on the currently available observational data it is not possible to unambiguously identify the preferred approach to cosmology.

   It has been phenomenologically shown that the tired light may be due to Mach effect, which may contribute dominantly to the cosmological redshift [10].  While the paper's assumption that observed redshift may be a combination of the expansion of the universe and tired light effect appears to be sound, it incorrectly divided the distance modulus of the light emitting source between the two components rather than keeping the proper distance of the source the same and dividing the redshift.  This was corrected in a subsequent paper [11] which showed that a hybrid of Einstein de Sitter cosmological model and the tired light (Mach effect) model gave an excellent fit to the SNe Ia data while at the same time providing analytically the deceleration parameter and the ratio of the contribution of the two models. Consequently, the new model was dubbed Einstein de Sitter Mach (EDSM) model.

Using Poisson's work on the motion of point particles in curved spacetime [12], Fischer [13] has shown analytically that gravitational back reaction may be responsible for the tired light phenomenon and could account for some or most of the observed redshift. His finding may also be related to Mach effect.

The EDSM model required a luminosity flux correction factor proportional to $1/\sqrt{1+z}$ that was left unexplained [11], $z$ being the redshift. This inspired us to look at the fundamentals of cosmological modeling and see if some of the assumptions need revisiting. Most cosmological models are based on one or more of the following assumptions:
1. Cosmological principle: The universe is homogeneous and isotropic – at large scale.
2. Adiabatic expansion: The energy does not enter or leave a volume of the universe.
3. Perfect fluid: The equation of state follows simple energy–pressure proportionately law.
4. Interaction free components: Fluid equation for each component is independent.

We believe that the adiabatic expansion of the universe is the weakest among all the above assumptions. After all Einstein's incorporation of the cosmological constant in his field equations in itself comprises a breach of adiabatic assumption. More recently, Komatsu and Kimura [14, 15] have suggested a non-adiabatic model. Their approach has been to modify the Friedmann and acceleration equations by adding extra terms and derive the continuity (fluid) equation from the first law of thermodynamics, assuming non-adiabatic expansion caused by the entropy and temperature on the horizon. The solution of the equations is thus based on multiple unknown parameters that need to be determined by fitting the SNe Ia data. We believe if the model is sound then we would not need any fitting parameter other than the Hubble constant.

## 2. Theory

The Friedmann equation, coupled with the fluid equation and the equation of state, provides the dynamics of the universe and thus the evolution of the scale factor $a$. It *does not* give the redshift $z$ directly. The redshift is taken to represent the expansion, and only the expansion, of the universe, and thus scale factor is considered to be directly observable through the relation $a = 1/(1+z)$. The relation ignores other causes that may contribute to the redshift. If the redshift is indeed contributed partially by other factors, such as by the Mach effect, then the scale factor determined by said equations will not equate to $1/(1+z)$. Unless the said equations are modified to take into account other factors, they cannot be considered to represent the cosmology correctly. Since energy density is common to all the three equations, and evolution of density is governed by the fluid equation, we will try to look at it with a magnifying glass.

The starting point for the fluid equation in cosmology is the first law of thermodynamics [1,16]:

$$dQ = dE + dW, \qquad (1)$$

where $dQ$ is the thermal energy transfer into the system, $dE$ is the change in the internal energy of the system, and $dW = PdV$ is the work done on the system having pressure $P$ to increase its volume by $dV$. Normally, $dQ$ is set to zero on the ground that the universe is perfectly homogeneous and that there can therefore be no bulk flow of thermal energy. However, if the energy loss of a particle, such as that of a photon through tired light phenomenon, is equally shared by all the particles of the universe (or by the 'fabric' of the universe) in the spirit of the Mach effect [17] then $dQ$ can be non-zero while conserving the homogeneity of the universe.

We will thus abandon the assumption that $dQ = 0$. The first law of thermodynamics for the expanding universe then yields:

$$\dot{E} + P\dot{V} = \dot{Q}. \qquad (2)$$

We now apply it to an expanding sphere of commoving radius $r_s$ and scale factor $a(t)$. Then the sphere volume $V(t) = \frac{4\pi}{3}r_s^3 a(t)^3$, and

$$\dot{V} = V\left(\frac{3\dot{a}}{a}\right). \qquad (3)$$

Since the internal energy of the sphere with energy density $\varepsilon(t)$ is $E(t) = \varepsilon(t)V(t)$, its rate of change may be written as

$$\dot{E} = V\dot{\varepsilon} + \dot{V}\varepsilon = V\left(\dot{\varepsilon} + \frac{3\dot{a}}{a}\varepsilon\right). \qquad (4)$$

If we assume the energy loss $\dot{Q}$ to be proportional to the internal energy $E$ of the sphere

$$\dot{Q} = -\beta E = -\beta \varepsilon V, \qquad (5)$$

where $\beta$ is the proportionality constant, then Equation (2) may be written as

$$\dot{\varepsilon} + \frac{3\dot{a}}{a}(\varepsilon + P) + \beta \varepsilon = 0, \qquad (6)$$

which is the new fluid equation for the expanding universe. Using the equation of state relation $P = w\varepsilon$, and rearranging Equation (6), we may write

$$\frac{d\varepsilon}{\varepsilon} + 3(1+w)\frac{da}{a} + \beta dt = 0. \qquad (7)$$

Assuming $w$ to be constant in the equation of state, this can be integrated to yield

$$\ln(\varepsilon) + 3(1+w)\ln(a) + \beta t + C = 0. \qquad (8)$$

Here $C$ is the integration constant. Now $t = t_0$ corresponds to the scale factor $a = 1$ and $\varepsilon = \varepsilon_0$, giving $C = -\ln(\varepsilon_0) - \beta t_0$. We may then write Equation (8) as

$$\varepsilon(a) = \varepsilon_0 a^{-3(1+w)} e^{\beta(t_0 - t)}. \qquad (9)$$

Let us now examine the simplest form of the Friedmann equation (single component, flat universe) with G as the gravitational constant. It may be written [16] as

$$\left(\frac{\dot{a}}{a}\right)^2 = \left(\frac{8\pi G \varepsilon}{3c^2}\right). \qquad (10)$$

Substituting $\varepsilon$ from Equation (9), we get

$$\dot{a}^2 = \left(\frac{8\pi G \varepsilon_0}{3c^2}\right) a^{-(1+3w)} e^{\beta(t_0 - t)}. \qquad (11)$$

Since $a_0 \equiv a(t_0) = 1$, it can be shown that it has the following solution [Appendix A, Equations (A1) to (A8)]:

$$a = a/a_0 = \left(\frac{1-e^{-\frac{\beta t}{2}}}{1-e^{-\frac{\beta t_0}{2}}}\right)^{\frac{2}{3+3w}}, \qquad (12)$$

$$\approx \left(\frac{t}{t_0}\right)^{\frac{2}{3+3w}} \left(1 + \frac{1}{4}\beta\left(\frac{2}{3+3w}\right)(t_0 - t) + O(\beta^2)\right). \qquad (13)$$

This reduces to the standard expression for the scale factor in adiabatic universe ($\beta = 0$). Since the Hubble parameter is defined as $H(t) = \dot{a}/a$, differentiating Equation (12) with respect to $t$ and rearranging, we get [Appendix A, Equations (A9) to (A??)]:

$$\frac{\dot{a}}{a} = \left(\frac{\beta}{3+3w}\right)\left(e^{\frac{\beta t}{2}} - 1\right)^{-1}, \text{ or} \qquad (14)$$

$$e^{\frac{\beta t}{2}} = 1 + \left(\frac{1}{H(t)}\right)\left(\frac{\beta}{3+3w}\right), \text{ or} \qquad (15)$$

$$\frac{\beta t}{2} = \ln\left(1 + \left(\frac{1}{H(t)}\right)\left(\frac{\beta}{3+3w}\right)\right), \text{ or} \qquad (16)$$

$$t_0 = \frac{2}{3+3w}\left(\frac{1}{H_0}\right) \text{ when } \beta \Rightarrow 0. \qquad (17)$$

Here Equations (15) and (16) can be used to determine the age of the universe in the non-adiabatic universe provided we know $\beta$. They reduce to Equation (17) in the limit of $\beta \Rightarrow 0$. It is the standard expression in adiabatic universe for the age of the universe in terms of the Hubble constant for a single component flat universe. We see from Equation (11) that at $t = t_0$, $\dot{a}(t_0) = \sqrt{\left(\frac{8\pi G\varepsilon_0}{3c^2}\right)}$. We can therefore write the expression for the age of the universe in terms the energy density as

$$t_{0,\beta} = \frac{2}{\beta}\ln\left(1 + \left(\frac{\beta}{2}\right)\left(\frac{1}{1+w}\right)\sqrt{\frac{c^2}{6\pi G\varepsilon_0}}\right), \text{ and} \qquad (18)$$

$$t_{0,0} = \left(\frac{1}{1+w}\right)\sqrt{\frac{c^2}{6\pi G\varepsilon_0}} \text{ when } \Rightarrow 0. \qquad (19)$$

Equation (18) is the expression for the age of the universe for the single component flat non-adiabatically expanding universe and Equation (19) is the standard expression for adiabatically expanding universe obtained in the limit of zero $\beta$. We need to know $\beta$ in order to get $t_0$ in the non-adiabatic universe.

If we like, we could resolve the Friedmann Equation (11) into adiabatic and non-adiabatic components:

$$\dot{a}^2 = \left(\frac{8\pi G\varepsilon_0}{3c^2}\right)a^{-(1+3w)}\left[1 + \beta(t_0 - t) + \frac{1}{2}\beta^2(t_0 - t)^2 \ldots\right]. \qquad (11')$$

Here 1st term in the square bracket is the adiabatic term that is used in most cosmological models and the remaining terms represent the non-adiabatic correction. The non-adiabatic correction is non-existent at $t = t_0$, i.e. $z = 0$, and negligible when $t$ is close to $t_0$, i.e. $z \ll 1$. Thus, we can resort to adiabatic universe as the boundary condition when finding certain analytical parameters and correlations. Since we know the analytically derived value of the deceleration parameter $q_0 = -0.4$ from the adiabatic EDSM model [11], let us first workout the expression for the same from its standard definition and see if $\beta$ can be expressed in terms of $q_0$.

$$q_0 \equiv -\left(\frac{\ddot{a}a}{\dot{a}^2}\right)_{t=t_0}. \tag{20}$$

Equation (14) may be differentiated and rearranged to obtain the expression for $q_0$ as follows.

$$\ddot{a}(t) = \left(\frac{\beta}{3+3w}\right)\left[\dot{a}(t)\left(e^{\frac{\beta t}{2}} - 1\right)^{-1} - a(t)\left(e^{\frac{\beta t}{2}} - 1\right)^{-2} e^{\frac{\beta t}{2}}\left(\frac{\beta}{2}\right)\right], \tag{21}$$

$$= \left(\frac{\beta}{3+3w}\right)\left(e^{\frac{\beta t}{2}} - 1\right)^{-1}\left[\dot{a}(t) - \left(\frac{\beta}{2}\right)a(t)\left(e^{\frac{\beta t}{2}} - 1\right)^{-1} e^{\frac{\beta t}{2}}\right], \tag{22}$$

$$= \left(\frac{\dot{a}(t)}{a(t)}\right)\dot{a}(t)\left[1 - \left(\frac{\beta}{2}\right)\left(\frac{a(t)}{\dot{a}(t)}\right)\left(e^{\frac{\beta t}{2}} - 1\right)^{-1} e^{\frac{\beta t}{2}}\right], \text{ or} \tag{23}$$

$$\frac{\ddot{a}(t)a(t)}{\dot{a}^2(t)} = 1 - \left(\frac{\beta}{2}\right)\left(\frac{a(t)}{\dot{a}(t)}\right)\left(e^{\frac{\beta t}{2}} - 1\right)^{-1} e^{\frac{\beta t}{2}}, \tag{24}$$

$$= 1 - \frac{3+3w}{2} e^{\frac{\beta t}{2}} \text{ from Equation (14), or} \tag{25}$$

$$q = -1 + \left(\frac{3(1+w)}{2}\right)e^{\frac{\beta t}{2}}, \text{ or } q_0 = -1 + \left(\frac{3(1+w)}{2}\right)e^{\frac{\beta t_0}{2}}. \tag{26}$$

For $q_0 = -0.4$ and $w = 0$ (i.e. matter only universe), Equation (26) yields $e^{\frac{\beta t_0}{2}} = 0.4$ or $\beta = -1.833/t_0$. Substituting these values in Equation (15) at $t = t_0$ yields $\beta = -1.8H_0$ and the age of the universe $t_0 = 1.02 H_0^{-1}$.

Up until now we have not used any observational data. In order to proceed further, we need to know the Hubble constant $H_0$. The observational data is usually provided in the form of distance modulus $\mu$ and the redshift $z$. In an expansion only model, we may write the distance modulus as [1,16]

$$\mu = 5\log(d_L) + 25, \text{ where} \tag{27}$$
$$d_L = (1+z)d_P, \text{ where} \tag{28}$$
$$d_P(t_0) = c\int_{t_e}^{t_0} \frac{dt}{a(t)}. \tag{29}$$

Here $d_L$ is the luminosity distance of the source emitting the photons at time $t_e$ whose redshift is being measured, and $d_P$ is the proper distance of the source in mega parsecs observed at time $t_0$. When all the redshift is allocated to the expansion of the universe, $1 + z = 1/a(t)$. We may then write Equation (29)

$$d_P(z) = c\int_{z(t_e)}^{z(t_0)} dz(1+z)/\left(\frac{dz}{dt}\right). \tag{30}$$

Equation (12) can now be used to determine $dz/dt$ for substitution in Equation (30). Since we are observing redshift in the matter dominated universe, we may simplify Equation (12) by taking $w = 0$, and rewrite it as

$$1 + z = \frac{1}{a} = \left(1 - e^{-\frac{\beta t_0}{2}}\right)^{\frac{2}{3}}\left(1 - e^{-\frac{\beta t}{2}}\right)^{-\frac{2}{3}}, \text{ or} \tag{31}$$

$$\frac{dz}{dt} = \left(\frac{\beta}{3}\right)(1+z)\left(1-e^{-\frac{\beta t}{2}}\right)^{-1}\left(-e^{-\frac{\beta t}{2}}\right). \tag{32}$$

We can use Equation (31) to express $\left(1-e^{-\frac{\beta t}{2}}\right)^{-1}$ and $\left(-e^{-\frac{\beta t}{2}}\right)$ in terms of $1+z$. By defining the constant term $\left(1-e^{-\beta t_0/2}\right) \equiv A$, we may write

$$\left(1-e^{-\frac{\beta t}{2}}\right) = A/(1+z)^{\frac{3}{2}}, \tag{33}$$

and rewrite Equation (32) as

$$\frac{dz}{dt} = \left(\frac{\beta}{3}\right)[(1+z)^{\frac{5}{2}}/A]\left[\frac{A}{(1+z)^{\frac{3}{2}}} - 1\right], \text{ or} \tag{34}$$

$$= \left(\frac{\beta}{3}\right)(1+z)\left[1 - \frac{(1+z)^{\frac{3}{2}}}{A}\right]. \tag{35}$$

Equation (30) may now be written

$$d_P(z) = -\left(\frac{3c}{\beta}\right)\int_0^z du \left[1 - \frac{(1+u)^{\frac{3}{2}}}{A}\right]^{-1}. \tag{36}$$

There is no simple analytical solution for the integral in Equation (36). Substituting $A \equiv (1-e^{-\beta t_0}) = -1.5$ and $\beta = -1.8H_0$ from above - Equation (26) and the paragraph following it, and defining $R_0 \equiv c/H_0$, we may write the distance modulus $\mu$ as

$$\mu = 5\log[\frac{R_0}{0.6}(1+z)\int_0^z du \left(1 + \left(\frac{2}{3}\right)(1+u)^{\frac{3}{2}}\right)^{-1}] + 25. \tag{37}$$

We can include Mach effect contribution to the redshift following the approach in an earlier paper [11] and recalculate the distance modulus $\mu$. Using subscript M for Mach effect and X for expansion effect and equating the proper distance expressions for the two, and since $1+z = (1+z_M)(1+z_X)$ and $R_0 z = R_M z_M = R_X z_M$, we may write

$$R_M \ln(1+z_M) = \left(\frac{R_X}{0.6}\right)\int_0^{z_X} du \left[1 + \frac{(1+u)^{\frac{3}{2}}}{1.5}\right]^{-1}, \text{ or} \tag{38}$$

$$\left(\frac{R_0 z}{z_M}\right)\ln(1+z_M) = \left(\frac{R_0 z}{0.6 z_X}\right)\int_0^{z_X} du \left[1 + \frac{(1+u)^{\frac{3}{2}}}{1.5}\right]^{-1}, \text{ or} \tag{39}$$

$$\left(\frac{R_0 z(1+z_X)}{z-z_X}\right)\ln((1+z)/(1+z_X)) = \left(\frac{R_0 z}{0.6 z_X}\right)\int_0^{z_X} du \left[1 + \frac{(1+u)^{\frac{3}{2}}}{1.5}\right]^{-1}. \tag{39'}$$

It is not possible to express analytically $z_X$ (or $z_M$) in terms of $z$ and write $\mu$ directly in terms of $z$. Nevertheless, Equation (39') can be numerically solved for $z_X$ for any value of $z$, and distance modulus calculated to include Mach effect as well as expansion effect using the expression

$$\mu = 5\log[\tfrac{R_0}{0.6}(\tfrac{z}{z_X})\int_0^{z_X} du \left(1 + \left(\tfrac{2}{3}\right)(1+u)^{\tfrac{3}{2}}\right)^{-1} \sqrt{(1+z_X)(1+z)}] + 25. \qquad (40)$$

As equality between the proper distances determined by the Mach effect and the expansion effect is already established by Equation (39), and since $1 + z = (1 + z_M)(1 + z_X)$, exactly the same result as from Equation (40) is obtained if we use the expression for $\mu$ as follows:

$$\mu = 5\log[R_0(\tfrac{z}{z_M})\ln(1+z_M)\sqrt{(1+z_X)(1+z)}] + 25, \text{ or} \qquad (41)$$

$$\mu = 5\log[R_0(\tfrac{z}{z_M})\ln(1+z_M)(1+z)/\sqrt{(1+z_M)}] + 25. \qquad (41')$$

We will now consider how various parameters compare between the adiabatic models and the non-adiabatic model developed here.

If we compare Equation (5) with the standard Mach effect photon energy loss equation $-\tfrac{dE}{dt} = H_0 E$ then for radiation energy $\beta = -H_0$ and the Mach effect redshift $z$ is given by $1 + z = \exp(H_0 d/c)$ [10] with $d = c(t_0 - t)$. And, since $w = 1/3$ for radiation, Equation (9) may be written for radiation as

$$\varepsilon_r(a) = \varepsilon_{r,0} a^{-4}(1+z). \qquad (42)$$

What is the scale factor here? In a standard adiabatically expanding universe, $\beta = 0$, and $a = 1/(1+z)$. However, in the EDSM model the redshift $z$ has two components, $z_X$ due to the expansion of the universe and $z_M$ due to the Mach effect, with $1 + z = (1 + z_X)(1 + z_M)$ [11]. We should therefore replace in Equation (42) $a$ with $a_X = 1/(1+z_X)$ and $z$ with $z_M$:

$$\varepsilon_r(a_X) = \varepsilon_{r,0} a_X^{-4}(1+z_M) = \varepsilon_{r,0} a_X^{-3}(1+z). \qquad (43)$$

Recalling that the standard expression for the radiation energy density evolution is given by

$$\varepsilon_r(a) = \varepsilon_{r,0} a^{-4} = \varepsilon_{r,0}(1+z)^4, \qquad (44)$$

we find that $\varepsilon_r(a_X) = \varepsilon_r(a)/(1+z_M)^3$, and is a fraction of the energy density for a given $z$ without the Mach effect.

Since Equation (9) is valid also for matter with $w = 0$, we get in the adiabatic universe with $\beta = 0$,

$$\varepsilon_m(a_X) = \varepsilon_{m,0} a_X^{-3}. \qquad (45)$$

Comparing it with Equation (43) we see that the ratio of the radiation density and mass density is proportional to $(1+z)$, the same as in the standard expansion models [1,16]. The ratio is inclusive of the factor $e^{-\beta(t_0-t)}$ in the non-adiabatic universe.

3. **Results**

The database used in this study is for 580 SNe Ia data points with redshifts $0.015 \leq z \leq 1.414$ as compiled in the Union2 $\mu, z$ database [18] updated to 2017.

We used Matlab curve fitting tool to fit the data using non-linear least square regression. To minimize the impact of large scatter of data points, we applied the 'Robust Bisquare' method in Matlab. This tool fits data by minimizing the summed square of the residuals, and reduces the weight of outliers using bi-square weights. This scheme minimizes a weighted sum of squares, where the weight given to each data point depends on how far the point is from the fitted line. Points farther from the line get reduced weight. Robust fitting with bisquare weight uses an iteratively reweighted least square algorithm. The Goodness of Fit in Matlab is given by parameters SSE (sum of squares due to errors, i.e. summed square of residuals) that is minimized in the fitting algorithm; R-Square that indicates the proportionate amount of variation in the response variable explained by the independent variable in the model (larger the R-squared, more the variability explained by the model); and RMSE (root mean square error, i.e. standard error of the regression)- closer the value to zero, better is the data fit.

Figure 1 shows the curves fitted to the data set using Equation (40) and Equation (37), and using standard ΛCDM model [11] for comparison. The expression used for ΛCDM model is

$$\mu = 5\log[R_0 \int_0^z du/\sqrt{\Omega_{m,0}(1+u)^3 + 1 - \Omega_{m,0}}] + 5\log(1+z) + 25. \qquad (46)$$

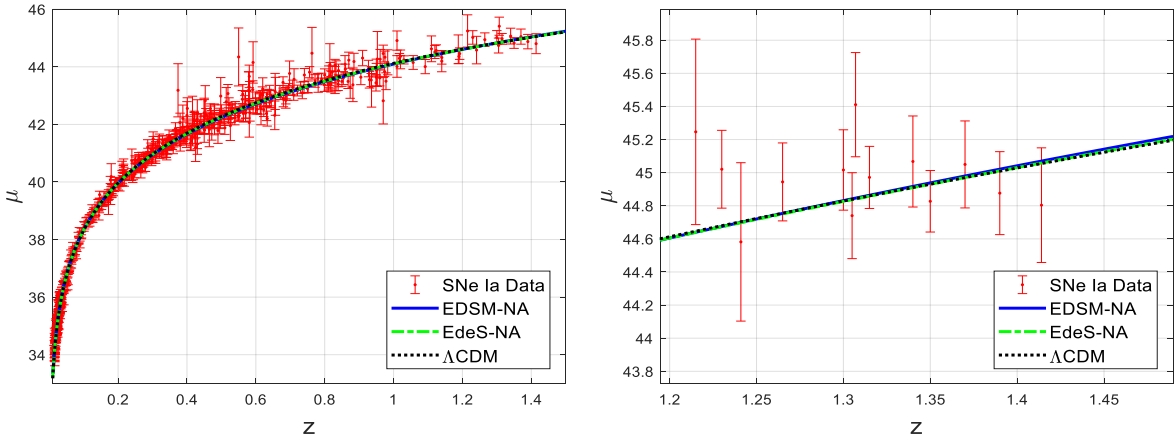

**Figure 1.** Fitted data curves for the three models in Table 1. The first one has been labeled as EDSM-NA since it is the non-adiabatic version of the EDSM model (flat, matter only, including Mach effect) in reference [11]. The second curve is labeled as EdeS-NA as it is the non-adiabatic version of the Einstein de Sitter model (flat, matter only universe). The third one is the curve for ΛCDM model. The left figure shows the complete fitted curves for the 580 points data set whereas the right figure is the zoom-in of the fit in the high $z$ region to enhance the difference among the fitted curves.

The first one has been labeled as EDSM-NA since it is the non-adiabatic version of the EDSM model (flat, matter only, including Mach effect) of reference [11]. The second curve is labeled as EdeS-NA as it is the non-adiabatic version of the Einstein de Sitter model (flat, matter only universe). The third one is the curve for standard ΛCDM model. There is no visible difference between the three curves when the full data fit curves are viewed in the left display of the figure, and only very slight visible difference when the zoomed-in right display at high $z$ is viewed. Corresponding goodness-of-fit numbers are presented in Table 1. The goodness-of-fit numbers differ only slightly.

Figure 2 depicts the evolution of dimensionless parameters: a) scale factor $a$, b) Hubble parameter $H/H_0$, c) and deceleration parameter $q$, against the dimensionless time $H_0(t - t_0)$ for the standard ΛCDM model and the non-adiabatic model developed here. While the shapes of the curves are different, the trends are similar. We notice that the ΛCDM curves are steeper in most of the plotted region.

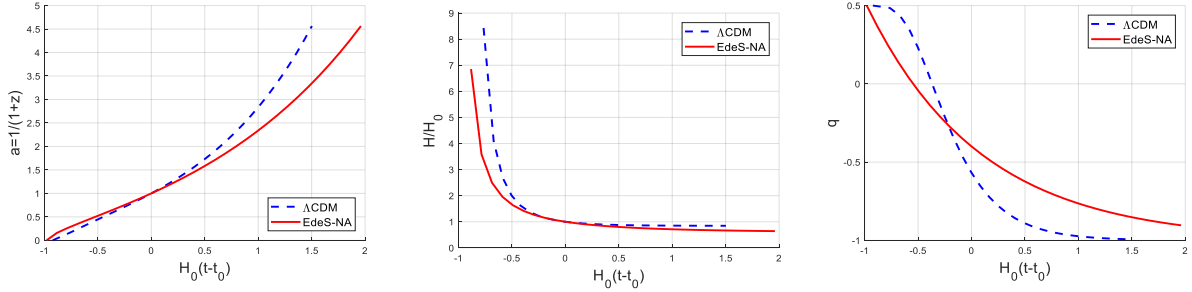

**Figure 2**. Evolution of dimensionless parameters - scale factor $a$ (left figure), Hubble parameter $H/H_0$ (middle figure), and deceleration parameter $q$ (right figure) – against dimensionless time $H_0(t - t_0)$ for the standard ΛCDM model and the Non-Adiabatic model EdeS-NA.

**Table 1.** Parameter and goodness-of-fit for the two models. $H_0$ is in km s$^{-1}$Mpc$^{-1}$. SSE stands for sum of squares due to errors and RMSE for root mean square error.

| Model | Parameter $H_0$ | 95% Confidence $H_0$ Low | $H_0$ High | Parameter $\Omega_{m,0}$ | 95% Confidence $\Omega_{m,0}$ Low | $\Omega_{m,0}$ High | Goodness of Fit SSE | R-Square | RMSE | Equation used |
|---|---|---|---|---|---|---|---|---|---|---|
| EDSM-NA | **68.28** | 68.81 | 67.75 | **None** | NA | NA | **24.58** | 0.9958 | 0.2060 | 40 |
| EdeS-NA | **69.01** | 69.54 | 68.48 | **None** | NA | NA | **24.10** | 0.9959 | 0.2040 | 37 |
| ΛCDM | **69.85** | 70.71 | 69.01 | **0.2877** | 0.2489 | 0.3266 | **24.35** | 0.9959 | 0.2053 | 46 |

In Figure 3 we have plotted inverse of the expansion scale factor $\frac{1}{a_X} = (1 + z_X)$ against the inverse of the standard scale factor $\frac{1}{a} = (1 + z)$. The curve can be approximated with a power law expression $y = 1.1345 x^{0.4714}$ except at rather low values. Also we have included a curve showing $a_X^{-3}(1 + z) = (1 + z_X)^3 (1 + z)$ against $a^{-4} = (1 + z)^4$ to show that the radiation energy density scaling is altered drastically by the inclusion of Mach effect. This curve may be approximated with a power law expression $y = 1.4604 x^{0.6035}$ except at small values.

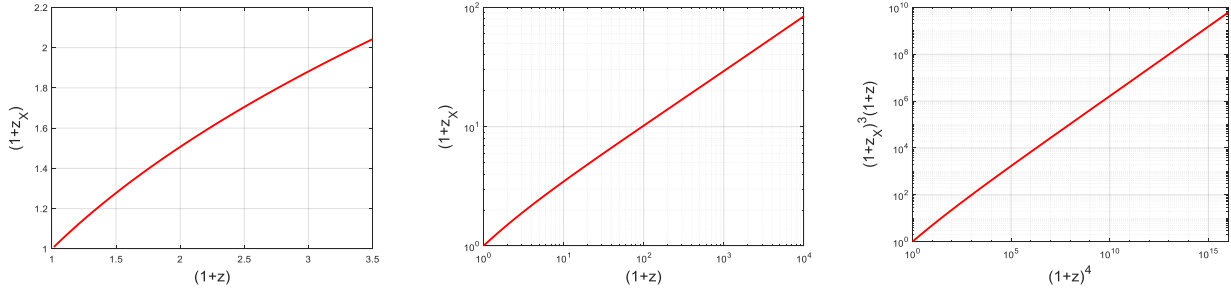

**Figure 3.** Evolution of inverse of scale factors and radiation density – Inverse expansion scale factor $1/a_X = (1 + z_X)$ against the inverse standard scale factor $1/a = (1 + z)$ for low values (left figure) and high values (middle figure); radiation energy density scaling (right figure) for the Non-Adiabatic model EDSM-NA .

## 4. Discussion

From Figure 1 and Table 1, it is difficult to state unarguably which model is better. Nevertheless, based on the fact that non-adiabatic models yield data fit using only one fit parameter, whereas the ΛCDM model requires two fit parameters, the preferred model would be one of the non-adiabatic models. And since the analytical value of the deceleration parameter used in this work is derived by equating the Mach proper distance and expansion proper distance [11], both the non-adiabatic models – EdeS-NA and EDSM-NA implicitly involve Mach effect. Our choice of the non-adiabatic model will thus be determined by studying which model fits other cosmological observations better.

It should be mentioned that non-adiabatic modeling has been tried by some cosmologist, most recently by Komatsu and Kimura [14, 15]. Their approach has been to modify the Friedmann and acceleration equations by adding extra terms and derive the continuity (fluid) equation from the first law of thermodynamics, assuming non-adiabatic expansion caused by the entropy and temperature on the horizon. The solution of the equations is thus based on multiple unknown parameters that need to be determined by fitting the SNe Ia data. Our approach here modifies only the fluid equation from the first law of thermodynamics on the assumption that the system energy gain or loss is proportional to the energy of the system – Equation (5). No adjustable parameters, other than the universal Hubble constant, are required to fit the data. The deceleration parameter that is needed in our non-adiabatic formulation is analytically obtained from the EDSM model [11]. It may therefore be concluded that the luminosity flux correction factor of reference [11] proportional to $1/\sqrt{1+z}$ is due to the non-adiabatic nature of the universe. Similarly, one could say that the cosmological constant Λ approximates the non-adiabatic nature of the universe when studied in an adiabatic approximation of the universe.

The lowest value of the Hubble constant in this work is obtained with the EDSM-NA model without compromising the goodness-of-fit. It is closer to the Hubble constant obtained from the cosmic microwave background (CMB) data, such as from Plank and WMAP space crafts, than from the ΛCDM model or the EdeS-NA model. At 68.28 km s$^{-1}$Mpc$^{-1}$ it is almost right at the weighted average of 68.1 km s$^{-1}$Mpc$^{-1}$ reported from WMAP and Planck data points [19]. However, it may just be a coincidence. As discussed by Bonnet-Bidaud [20], the origin of CMB is not fully settled as yet.

Another important thing to note is that the ratio of Mach and expansion contribution to the redshift has now changed. In reference [11] expansion contribution at the current epoch ($t = t_0$ or $z = 0$) was only 40%. If we expand Equation (39) in the limit of very small $z$, we can see that $z_M = 0.6 z_X$ and since $z = z_M + z_X$ in this limit, we find that $z_X$ is 62.5% of $z$. The luminosity flux correction factor can be considered as responsible for this discrepancy. This means that 62.5% of Hubble constant $H_0$ is due to the expansion of the universe and remaining due to Mach effect. All the expansion related cosmological parameters should thus be determined using $H_{0X} = 43$ km s$^{-1}$Mpc$^{-1}$. For example the age of the universe would be $t_0 = \frac{1.02}{H_{0X}} = 1.632\ H_0 \approx 23$ Gyr in the EDSM-NA model against $t_0 = \frac{1.02}{H_0} \approx 14.5$ Gyr in EdeS-NA model.

It should be emphasized that the main merit of the model presented here is that it can fit the data with a single parameter, the Hubble constant $H_0$. There have been several models developed in the past, such as based on the modified tired light approach in plasma cosmology by Lorenzo Zaninetti [21], that can give excellent fit to the data with one additional parameter which has to be determined by fitting the data.

## 5. Conclusion

The cosmological model presented in this communication is based on relaxing the assumption that universe dynamics is adiabatic within the confines of the cosmological principle. The fact that a single parameter yields a better fit to the SNe Ia data using the non-adiabatic model presented here than the two parameter fit of the same data using ΛCDM model establishes the superiority of the new model. The Hubble constant obtained by the two models is almost the same, in fact the non-adiabatic Mach-expansion hybrid model EDSM-NA gives a lower value, $H_0 = 68.28\ (\pm 0.53)$ km s$^{-1}$Mpc$^{-1}$, very close to 68.1 km s$^{-1}$Mpc$^{-1}$ sought by cosmic microwave background data from Plank and WMAP space crafts. It may therefore be possible to dispense with the cosmological constant after all, and corresponding perpetually elusive dark energy, in the spirit of Einstein who always wanted to correct his greatest mistake!

## Appendix A

In this Appendix, our objective is to show how to obtain Equations (12) and (14) from Equation (11).

Equation (11) may be rewritten as

$$\dot{a} = \left(\frac{8\pi G \varepsilon_0}{3c^2}\right)^{\frac{1}{2}} e^{\frac{\beta t_0}{2}} a^{-\frac{(1+3w)}{2}} e^{\frac{-\beta t}{2}}. \tag{A1}$$

Substituting temporarily $A = \left(\frac{8\pi G \varepsilon_0}{3c^2}\right)^{\frac{1}{2}} e^{\frac{\beta t_0}{2}}$ and $B = 1 + 3w$, we may write Equation (A1) as

$$\dot{a} = A a^{-\frac{B}{2}} e^{\frac{-\beta t}{2}}. \tag{A2}$$

The solution of this equation from the free online solver Wolfram Alpha (http://www.wolframalpha.com) is

$$a(t) = 4^{-\frac{1}{B+2}} \left[(-B-2)\left(\frac{2Ae^{-\frac{\beta t}{2}}}{\beta} + c1\right)\right]^{\frac{2}{B+2}}. \tag{A3}$$

Here $c1$ is the integration constant that needs to be determined from the boundary condition; it should reduce to the standard expression for $a(t)$ [16] in the non-adiabatic universe when $\beta = 0$. Since the scale factor $a(t_0) \equiv 1$, dividing Equation (A3) by the same by setting $t = t_0$, we get

$$a(t) = \frac{a(t)}{a(t_0)} = \left[\left(\frac{2Ae^{-\frac{\beta t}{2}}}{\beta} + c1\right)\right]^{\frac{2}{B+2}} / \left[\left(\frac{2Ae^{-\frac{\beta t_0}{2}}}{\beta} + c1\right)\right]^{\frac{2}{B+2}}, \tag{A4}$$

$$= \left[\left(e^{-\frac{\beta t}{2}} + \beta c1/2A\right)\right]^{\frac{2}{B+2}} / \left[\left(e^{-\frac{\beta t_0}{2}} + \beta c1/2A\right)\right]^{\frac{2}{B+2}}, \tag{A5}$$

$$= \left[\left(1 - \frac{\beta t}{2} + \ldots + \beta c1/2A\right) / \left(1 - \frac{\beta t_0}{2} + \ldots + \beta c1/2A\right)\right]^{\frac{2}{3+3w}}, \tag{A6}$$

where we have used series expansion for the exponential function and retained only first two term in the numerator as well as denominator in order to take the limit $\beta \Rightarrow 0$. This must reduce to the standard expression for single component flat universe, i.e. to $\left(\frac{t}{t_0}\right)^{\frac{2}{3+3w}}$ [16]. With this boundary condition, we see that we must have $\frac{\beta c1}{2A} = -1$. We may now write

$$a(t) = \left[\left(e^{-\frac{\beta t}{2}} - 1\right) / \left(e^{-\frac{\beta t_0}{2}} - 1\right)\right]^{\frac{2}{3+3w}}, \tag{A7}$$

$$= \left[\left(1 - e^{-\frac{\beta t}{2}}\right) / \left(1 - e^{-\frac{\beta t_0}{2}}\right)\right]^{\frac{2}{3+3w}}. \tag{A8}$$

Equation (A8) is the same as Equation (12). Taking time derivative of this equation, we get

$$\dot{a}(t) = \frac{2}{3+3w} \left[\frac{\left(1-e^{-\frac{\beta t}{2}}\right)}{\left(1-e^{-\frac{\beta t_0}{2}}\right)}\right]^{\frac{2}{3+3w}} \left(1 - e^{-\frac{\beta t}{2}}\right)^{-1} \left(-e^{-\frac{\beta t}{2}}\right)\left(-\frac{\beta}{2}\right), \tag{A9}$$

$$= a(t)\left(\frac{2}{3+3w}\right)\left(1 - e^{-\frac{\beta t}{2}}\right)^{-1} \left(\frac{\beta}{2}\right) e^{-\frac{\beta t}{2}}, \text{ or} \tag{A10}$$

$$\frac{\dot{a}(t)}{a(t)} = \left(\frac{\beta}{3+3w}\right)\left(e^{\frac{\beta t}{2}} - 1\right)^{-1}. \tag{A11}$$

This expression is the same as Equation (14).